\documentclass[notitlepage,11pt]{article}
\usepackage{epsfig}
\usepackage{fullpage}
\usepackage{subfigure}
\usepackage[sort&compress,numbers]{natbib}
\usepackage{wrapfig}

\begin{document}

\begin{center}
{\Large Large Sample Superradiance and Fault-Tolerant Quantum Computation}

\vspace{0.2cm}

{\large D. D. Yavuz and B. Lemberger}

\vspace{0.2cm}

{\it Department of Physics, 1150 University Avenue,
University of Wisconsin at Madison, Madison, WI, 53706}

\end{center}

\vspace{1cm}

\noindent {\it \textbf{Abstract:}} We quantitatively analyze superradiance (collective emission) in a three-dimensional array of qubits without imposing any restrictions on the size of the sample. We show that even when the spacing between the qubits become arbitrarily large, superradiance produces an error rate on each qubit that scales with the total number of qubits. This is because the sum of the norms of the effective Hamiltonians that decoheres each qubit scales with the total number of qubits and is, therefore, unbounded. In three spatial dimensions, the sum of the norms scales as $N^{2/3}$ where $N$ is the total number of qubits in the computer. Because the sum of the Hamiltonian norms are unbounded, the introduced errors are outside the applicability of the threshold theorem. 

\newpage

Quantum computers utilize the exponentially large Hilbert space of identical two-level systems (qubits) and quantum entanglement to possibly solve certain problems much faster than any foreseeable classical computer \cite{nielsen,bennett,shor,jozsa,grover,abrams}. It is now well-understood that scalable computation requires error correction and fault tolerance, and quantum computers are no exception. Similar to classical computers, future quantum computers will almost certainly employ codes to detect and correct errors at various stages of the computation. It is not a priori obvious that traditional error correction ideas would extend to quantum computers. Over the last two decades, a growing body of literature has shown that error correction and fault-tolerant operation is indeed possible for quantum computers \cite{shor2,steane,aharonov}. One of the most important achievements in the field has been the discovery of the threshold theorem \cite{terhal,aharonov2,preskill,preskill2}. This theorem argues that  if the quantum gates are constructed with a fidelity better than a certain threshold, then arbitrarily long quantum operations are, in principle, possible. The error threshold is typically established to be $\eta \sim 10^{-4}$, but can be as high as  $\eta \sim 10^{-2}$ for surface codes \cite{fowler}. The threshold theorem has been a main driving force in the field. Many experimental implementations of quantum computing have been aiming to demonstrate gate fidelities better than the mentioned threshold, in order to demonstrate the feasibility of their approach. 

Although the threshold theorem is a remarkable achievement, it has a number of weaknesses. The theorem works under certain assumptions regarding the properties of the noise that affects the quantum computer. One of the key requirements is that sum of the norms of the Hamiltonians that couples each qubit to the bath must be bounded. It is now well-understood that this assumption is not valid for certain models of environment-qubit coupling, especially when the bath is bosonic in nature. In such cases, it has been discussed that the threshold theorem becomes extremely sensitive to the high frequency spectrum of the bath operators \cite{preskill}.  A number of authors have also criticized the threshold theorem using more general arguments \cite{staudt,dyakonov,alicki,kalai}. Despite these weaknesses, the threshold theorem has generally been regarded to cover all reasonable models of noise sources. In this letter, we extend our recent result \cite{yavuz} to large samples and discuss a clear physical mechanism which produces errors on the qubits that are beyond the applicability of the threshold theorem. We analyze superradiant emission between the qubit levels resulting from the interaction of the qubits with a common radiation bath in free space.  We show that the sum of the norms of the effective Hamiltonians that decoheres each qubit scales with the total number of qubits and is, therefore, unbounded. As a result, there is an error rate on each qubit that scales with the total number of qubits in the computer. Because the error rate scales with the total number of qubits, it cannot be assumed lower than a certain threshold; this is precisely how the assumptions of the threshold theorem are violated.  To our knowledge, this is the first time a clear physical noise source is identified that produces errors beyond the applicability of the threshold theorem. Below we discuss radiatively coupled qubits in free space, so our results are particularly relevant for neutral atom- \cite{saffman,grangier} and trapped-ion-based \cite{cirac,monroe} quantum computation. However, our results will likely be applicable to other physical systems, since a source of collective emission can be found in most situations, for example, phonons for solid-state-based approaches \cite{loss,clarke}.

\begin{figure}[tbh]
\vspace{-0cm}
\begin{center}
\includegraphics[width=15cm]{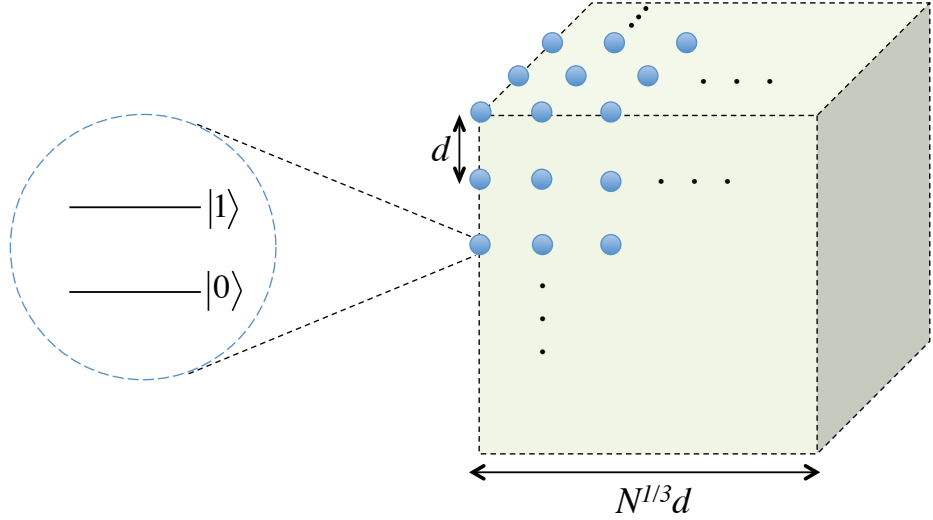}
\vspace{-1cm} \caption{\label{cube} \small We analyze an $N$ qubit quantum computer in a three dimensional geometry without any restrictions on the total size of the sample: i.e, the spacing between the qubits may be much larger than the radiation wavelength, $ d >> \lambda_a$. }
\end{center}
\vspace{-0.3cm}
\end{figure}

Since the seminal paper by Dicke \cite{dicke}, the superradiance problem has been analyzed by a large number of authors and this problem continues to be relevant for a wide range of physical systems \cite{haroche,yelin1,yelin2,francis}. As shown in Fig.~\ref{cube}, we consider $N$ two-level atoms, each with levels $|0\rangle$ and $|1\rangle$, in a three dimensional cube geometry. We will denote each individual qubit with the index $j$.  We will consider a continuum of radiation modes with annihilation and creation operators $\hat{a}_{k\epsilon}$ and $\hat{a}_{k\epsilon}^\dag$ respectively. These operators act on the mode of the field with wave-vector $k$ and polarization $\epsilon$. The total Hamiltonian for the system is:
\begin{eqnarray}
\hat{H}_{total}=\sum_{k,\epsilon}\hbar \nu_{k \epsilon} \left( \hat{a}_{k \epsilon}^\dag \hat{a}_{k \epsilon}+\frac{1}{2} \right) + \sum_{j}\hat{H}^j \quad , 
\end{eqnarray}

\noindent where
\begin{eqnarray}
\hat{H}^j & = &\frac{1}{2}\hbar \omega_a \hat{\sigma}_z^j-\sum_{k, \epsilon}\hbar g_{k \epsilon} \left[ \hat{a}_{k \epsilon} \exp{(i \vec{k} \cdot \vec{r}_j )} \hat{\sigma}_+^j+\hat{a}_{k \epsilon}^\dag \exp{(- i \vec{k} \cdot \vec{r}_j ) }\hat{\sigma}_{-}^j \right] \quad , \nonumber \\
\hat{\sigma}_z^j & =& |1\rangle^j \hspace{0.1cm} {^j}\langle 1|-|0\rangle^j \hspace{0.1cm} {^j}\langle 0| \quad , \nonumber \\
\hat{\sigma}_+^j & =& |1\rangle^j \hspace{0.1cm} {^j}\langle 0| \quad , \nonumber \\
\hat{\sigma}_{-}^j & =& |0\rangle^j \hspace{0.1cm} {^j}\langle 1| \quad .
\end{eqnarray}

\noindent Here, $\vec{r}_j$ is the position of the $j$'th atom. The energies of the qubit states $|0\rangle$ and $|1\rangle$ are taken to be $-\frac{1}{2} \hbar \omega_a$ and $\frac{1}{2} \hbar \omega_a$, respectively. We take the initial state of the atomic system to be an arbitrary (in general entangled) superposition state and assume zero photons in each field mode $k \epsilon$. The initial state of the combined atom-radiation field system can be written as:
\begin{eqnarray}
|\psi (t=0)\rangle = \sum_{q=0}^{2^N-1}c_q |q \rangle \otimes | 0 \rangle \quad . 
\end{eqnarray}

\noindent Here, the index $q$ sums over all $2^N$ possibilities for the collective atomic states and $0$ refers to having zero photons in all radiation modes. Using the total Hamiltonian of Eqs.~(1) and (2), and the initial state of Eq.~(3), we study the problem in the Schr\"odinger picture. Superradiance in large samples is known to be difficult to analyze quantitatively.  Motivated by the ideas that are discussed in Ref.~\cite{haroche}, we have developed a model that takes into account the collective Lamb shift and also light propagation effects between different regions of the sample. In this model, we view the problem as a convolution of the single-atom radiative decay (which is the well-known Wigner-Weisskopf theory \cite{yamamoto}) and the atom distribution diffraction function. The details of this model will be discussed in a future publication \cite{ben}. The end result is a set of coupled integro-differential equations for the probability amplitudes of Eq.~(3):
\begin{eqnarray}
\frac{dc_{q}}{dt} &  = & -\left( \frac{3}{8 \pi} \right) \frac{\Gamma}{2} \sum_{q'}  (1-\cos^2\theta_{jj'}) \int_0^t \exp{\left[ i \omega_a (t-\tau) \right] G(t-\tau) }  c_{q'}(\tau) d\tau   \quad . 
\end{eqnarray}

\noindent where  
\begin{eqnarray}
G(t-\tau) \equiv   \frac{2\pi}{(r_{jj'}/c)} {\mathrm{Box}}\left[ \frac{t-\tau}{(r_{jj'}/c)} \right] - i \frac{2}{(r_{jj'}/c)}  \ln \left[\frac{(r_{jj'}/c) + (t-\tau)}{\vert  (r_{jj'}/c) - (t-\tau) \vert } \right] \quad .
\end{eqnarray}

\noindent Here, the quantity $\Gamma$ is the single-atom decay rate and the summation $\sum_{q'}$ is over all states $|q'\rangle$ that either equals $|q\rangle$ or differs from $|q\rangle$ only in the states of two atoms. The indices of these two atoms are labeled by $j$ and $j'$. One of these atoms has changed its internal state from  $|1 \rangle$ to $|0 \rangle$   whereas the other one has changed it from  $|0 \rangle$ to $ |1 \rangle$. $r_{jj'}$ is the distance between the two atoms and the quantity $\theta_{jj'}$ is the angle between the atomic dipole moment vector and the separation vector $\vec{r}_{jj'}$. In Eq.~(5), the Box function equals one when its argument is between 0 and 1, and equals zero otherwise. Eqs.~(4) are a set of coupled integro-differential equations for $2^N$ coefficients and are in general very difficult to solve. We next simplify the problem by employing the Markov approximation, assume $c_{q'}(\tau) \approx c_{q'}(t)$, and take this quantity outside the integral. We have verified that this is a very good approximation at the initial stages of decay, by exactly solving Eqs.~(4) in a small subset of the Hilbert space. With this approximation, Eqs.~(4) reduce to a linear set of equations and  can be described by a time-dependent effective Hamiltonian, $\hat{H}^{eff}$. In frequency units, this effective Hamiltonian is:
\begin{eqnarray}
i \frac{dc_{q}}{dt} =\sum_{q'} \hat{H}^{eff}_{qq'}(t) c_{q'} \quad ,
\end{eqnarray}

\noindent and has the following matrix elements
\begin{eqnarray}
\hat{H}^{eff}_{qq'}(t)=-i \left( \frac{3}{8 \pi} \right) \frac{\Gamma}{2}   (1-\cos^2\theta_{jj'}) \int_0^t \exp{\left[ i \omega_a (t-\tau) \right] G(t-\tau) d \tau } \quad , 
\end{eqnarray}

\noindent when $|q\rangle$ differs from $|q'\rangle$ by raising one qubit and lowering another ($j$ and $j'$ label these raised and lowered qubits and $j=j'$ if $|q\rangle = |q'\rangle$). $\hat{H}^{eff}_{qq'}=0$ if $|q\rangle$ and $|q'\rangle$ do not have this relationship. We note that $\hat{H}^{eff}$ is not Hermitian since it incorporates radiative decay as well as the collective Lamb shift. We next discuss the error rate on each qubit and how the effective Hamiltonian of Eq.~(7) violates the threshold theorem. An important observation is that, $\hat{H}^{eff}$ can be written as a sum of two qubit interactions only:
\begin{eqnarray}
\hat{H}^{eff} = \sum_j \sum_{j'} \hat{H}^{jj'} \quad .
\end{eqnarray}

\noindent Here the sum is over all pairs of qubits and operators $\hat{H}^{jj'}$ act nontrivially only on the qubits with indices $j$ and $j'$. The matrix elements of $\hat{H}^{jj'}$ can be found by using the elements of the effective Hamiltonian of Eq.~(7). For each qubit $j$, the threshold theorem requires \cite{preskill,aharonov2,preskill2}
\begin{eqnarray}
\sum_{j'} ||\hat{H}^{jj'}|| t_0 < \eta ,
\end{eqnarray}

\begin{figure}[tbh]
\vspace{-0cm}
\begin{center}
\includegraphics[width=11cm]{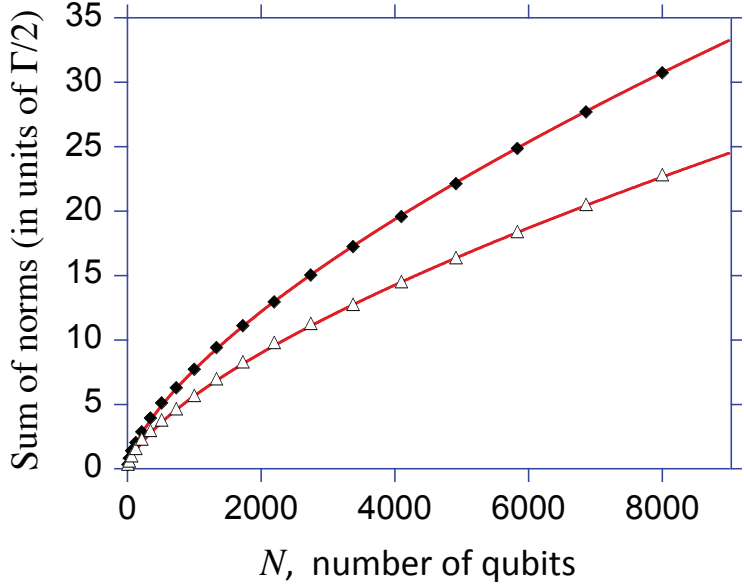}
\vspace{-0.6cm} \caption{\label{norm} \small  The sum of the Hamiltonian norms, $\sum_{j'} ||\hat{H}^{jj'}||$, as the total number of qubits is varied from 8 to 8000. The diamond data points are calculated by evaluating the effective Hamiltonian of Eq.~(7) at long time scales, $ t>> r_{jj'}/c$. The triangles are calculated using the Kurizki-Molmer (KM) Hamiltonian of Eq.~(10). The agreement between our model and the KM model is reasonable. Both models fit very well to $N^{2/3}$ power law (solid lines).  }
\end{center}
\vspace{-0.3cm}
\end{figure}

\noindent where $||\hat{H}^{jj'}||$ denotes the sup operator norm, $t_0$ is the time required for a quantum gate, and $\eta$ is the error threshold. Equation~(9) requires the sum of operator norms to be bounded, $\sum_{j'} ||\hat{H}^{jj'}|| < \infty$.  In Fig.~\ref{norm}, we numerically calculate  $\sum_{j'} ||\hat{H}^{jj'}||$ as the total number of qubits is varied from 8 (an array of $2 \times 2 \times 2$) to 8000 ($20 \times 20 \times 20$). Here, to calculate the norms, we evaluate the matrix elements of Eq.~(7) at long time scales, after full correlations have built-up in the system, $ t>> r_{jj'}/c$. In this numerical example we take the wavelength of the radiation to be $\lambda_a=9$~cm and the spacing between the atoms to be $d=20$~cm. We assume the direction of the atomic dipoles to be aligned with one of the axis of the cube (we have checked that the results remain very similar for any direction of the atomic dipole). The numerical data points fit very well to an $N^{2/3}$ power law which demonstrates $\sum_{j'} ||\hat{H}^{jj'}||$ to be unbounded. Physically, the reason for the $N^{2/3}$ dependence is that the interaction between the qubits due to superradiance decays slowly as a function  of the distance, only as $\sim 1/r$. In the three-dimensional cube geometry of  Fig.~1, the distance between the qubits scale as $r \sim N^{1/3} d$. There are $N$ terms in the sum of operator norms, $\sum_{j'} ||\hat{H}^{jj'}||$, resulting in the $N^{2/3}$ dependence.  As the qubit separation, $d$, is further increased, the sum of the norms plotted in Fig.~\ref{norm} would drop as $\sim 1/d$. However, the $N^{2/3}$ scaling with the total number of qubits would remain for arbitrarily large $d$.  In a two-dimensional geometry for the qubit array, the scaling would be $N^{1/2}$. We will discuss our two-dimensional simulations in detail elsewhere \cite{ben}. The calculation of Fig.~2 is performed for an atom at the corner of the cube; all of the other atoms in the array show a very similar behavior. 

We have also performed this calculation using the large-sample superradiance model which has recently been discussed by Kurizki {\it et al.} \cite{kurizki} and Molmer and colleagues \cite{molmer} [we will refer to this model as Kurizki-Molmer (KM) model]. This model ignores collective Lamb shift and makes approximations that are very similar to the single-atom Wigner-Weisskopf theory. In this model, the matrix elements of the effective Hamiltonian are given by:
\begin{eqnarray}
\hat{H}^{eff, KM}_{qq'}=-i \left( \frac{3}{8 \pi} \right) \frac{\Gamma}{2}  \left[ 4 \pi  (1-\cos^2\theta_{jj'}) \frac{\sin k_a r_{jj'}}{k_a r_{jj'}} 
+ 4 \pi (1-3\cos^2\theta_{jj'})  \left(\frac{\cos k_a r_{jj'}}{(k_a r_{jj'})^2} -\frac{\sin k_a r_{jj'}}{(k_a r_{jj'})^3} \right)   \right] \quad . 
\end{eqnarray}

\noindent Similar to Eq.~(7), Eq.~(10) holds when $|q\rangle$ and $|q'\rangle$ differ only in two qubits and $\hat{H}^{eff, KM}_{qq'}=0$ otherwise. The KM model is easier to calculate numerically since each matrix element can be algebraically evaluated (instead of computing an integral). As shown in Fig.~2, the agreement between the two models is reasonable and both models show a clear $N^{2/3}$ dependence of the sum of the norms. 

\begin{figure}[h]
\vspace{-0.8cm}
\begin{center}
\includegraphics[width=9cm]{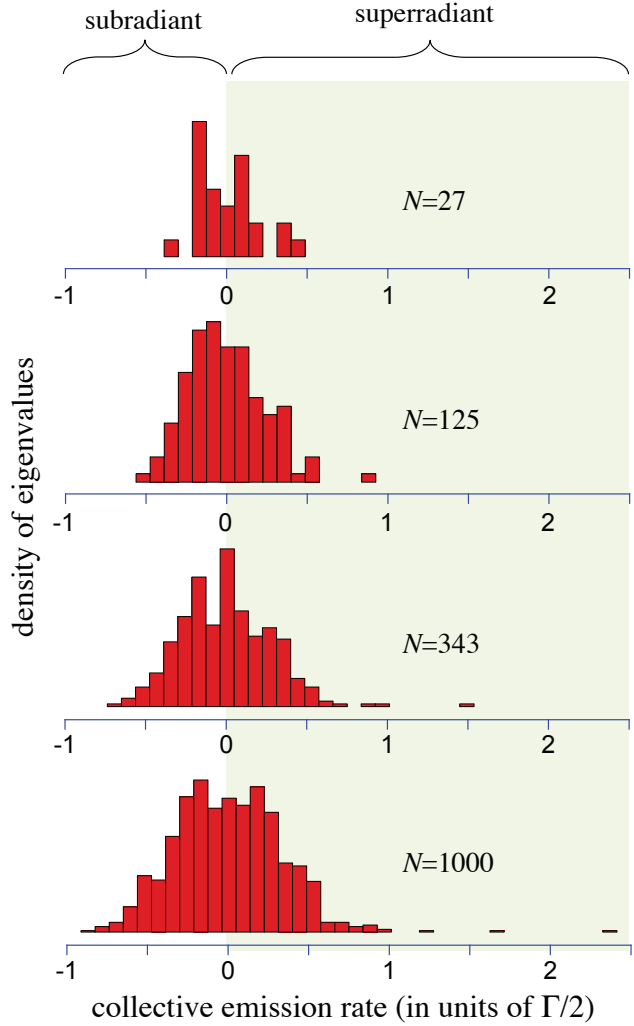}
\vspace{-1cm} \caption{\label{histograms} \small The histograms for the imaginary parts of the eigenvalues (i.e, the collective decay rates) for single atom excited states as the total number of qubits is varied from $N=27$ to $N=1000$. The histograms clearly show that there are states whose collective decay rates scale with the total number of qubits in the computer.  }
\end{center}
\vspace{-0.5cm}
\end{figure}

We note that the decomposition of $\hat{H}^{eff}$ as a sum of two-qubit interactions [Eq.~(8)] is the most physical one, but mathematically speaking this decomposition is not unique. $\hat{H}^{eff}$ can be written as a sum of 1, 2, ..., $k$ qubit interaction terms where $k \leq N$.  We have proven that, regardless of the choice of decomposition, either for some value of $k$, $\eta^{(k)}_1$ (as defined in Ref.~\cite{preskill2} - in the notation of Ref.~\cite{preskill2} , the $\eta$ in Eq.~(9) corresponds to $\eta^{(2)}_1$) diverges, or else the $\eta^{(k)}_1$ do not decay exponentially with $k$.  Either conclusion renders the threshold theorem inapplicable. We will discuss this proof in detail elsewhere \cite{ben}.  

To get physical insight into these results, we next discuss superradiance for states that have only one atom in the excited level $|1\rangle$. For these single-atom excited states, the dimension of the Hilbert space is $N\times N$ and the eigenvalues and eigenvectors of $\hat{H}^{eff}$ can be numerically evaluated. Figure~\ref{histograms} shows the histograms for the imaginary parts of the eigenvalues, which are the collective decay rates, as the number of qubits is varied from $N=27$ ($3 \times 3 \times 3$) to $N=1000$ ($10 \times 10 \times 10$). The histograms are calculated for parameters identical to those of Fig.~\ref{norm} with $\lambda=9$~cm and $d=20$~cm. In these histograms, we subtract the single atom decay rate of $\Gamma/2$ from the eigenvalues so that collective rates that are greater than zero correspond to superradiant states. From these histograms, we observe that there are states whose collective decay rates grow with $N$. The distributions remain roughly symmetric around zero, which means that about half of the states are superradiant whereas the other half are subradiant. The width of the distributions also grow with $N$, but very slowly. In these calculations, we could not locate a state with an eigenvalue of exactly zero, which means that there is not a state that does not decohere. This is consistent with the recent results of Whaley and colleagues who discussed that decoherence free subspaces do not exist for extended systems \cite{whaley}. 

\begin{figure}[h]
\vspace{-0cm}
\begin{center}
\includegraphics[width=16cm]{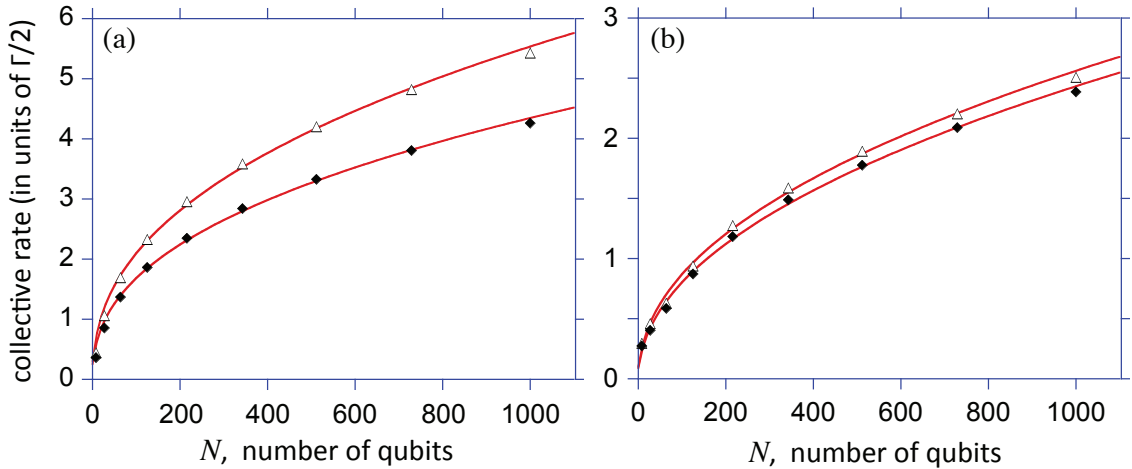}
\vspace{-0.3cm} \caption{\label{rate} \small  The largest collective rate for single-atom excited states as the number of qubits is varied from $N=8$ to $N=1000$ for (a) $d=15$~cm and (b) $d=20$~cm (the wavelength is $\lambda_a=9$~cm). The diamond data points are calculated using the effective Hamiltonian of Eq.~(7) and the triangles are calculated using the KM model of Eq.~(10). The solid lines are power-law fits to the data demonstrating $N^{0.41}$ scaling for (a) and $N^{0.48}$ for (b), for the diamond data points.  }
\end{center}
\vspace{-0.3cm}
\end{figure}

In Fig.~\ref{rate}, we plot the largest collective rate (i.e., the eigenvalue with the highest imaginary value from the histograms of Fig.~\ref{histograms}) as the number of qubits in the computer is varied from $N=8$ to $N=1000$. We perform these calculations for two different inter-qubit spacings, (a) $d=15$~cm and (b) $d=20$~cm (the wavelength remains $\lambda_a=9$~cm), both using the effective Hamiltonian of Eq.~(7) (diamonds) and also using the KM model of Eq.~(10) (triangles). We again observe reasonable agreement between the two models. The solid lines are power-law fits to the data. For these two calculations, the best fits to our model give $N^{0.41}$ ($d=15$~cm) and $N^{0.48}$ ($d=20$~cm) dependence of the largest collective rate on the number of qubits. We have performed this calculation for a wide range of parameters and have observed this dependence to vary between $N^{0.35}$ to $N^{0.5}$. These calculations show that even for single-atom excited states, there are states whose collective rates scale with the number of qubits. As a result, when the system is in one of these states, the errors introduced would also scale with the total number of qubits. 

In conclusion, we have discussed superradiance as a noise source which produces errors outside the applicability of the threshold theorem. There are many open questions and possible future directions: (i) Can noise due to superradiance be suppressed by modifying the vacuum modes, for example using a high-finesse cavity? (ii) Is scalable quantum computing possible while the system remains in subradiant states only, i.e, in the lower half of the histograms of Fig.~\ref{histograms}? (iii) Can the threshold theorem be extended to cover the superradiance noise that is discussed in this letter? If not, then what are the implications for scalable quantum computing? 

We thank Mark Saffman and Thad Walker for many helpful discussions and acknowledge support from the University of Wisconsin-Madison.

\bibliographystyle{plainnat}

\end{document}